\documentclass[oldversion,letter]{aa} 

\usepackage{natbib}

  \usepackage{subfigure}

\usepackage{graphicx}

\usepackage{txfonts}

\begin{document}

   \title{Detection of orbital parameter changes in the TrES-2 exoplanet ?}

   \author{D. Mislis          \and
          J.H.M.M. Schmitt          }

\institute{Hamburger Sternwarte, Gojenbergsweg 112, D-21029 Hamburg \\
              \email{mdimitri@hs.uni-hamburg.de}
             }

\date{Accepted : 18 May 2009}



\abstract
{We report a possible change in the orbit parameters of the TrES-2 exoplanet. With a period of 2.470621 days, the TrES-2 exoplanet exhibits almost "grazing" transits 110.4 minutes duration as measured in 2006 by Holman and collaborators. We observed two transits of TrES-2 in 2008 using the 1.2m Oskar-L\"{u}hning telescope (OLT) of Hamburg observatory employing CCD photometry in an i-band and a near to R-band filter. A careful lightcurve analysis including a re-analysis of the 2006 observations shows that the current transit duration 
has shortened since 2006 by $\approx 3.16$ minutes. Although the new observations were taken in a different filter we
argue that the observed change in  transit duration time cannot be attributed
to the treatment of limb darkening.
If we assume the stellar and planetary radii to be constant, a change in orbit inclination is the most likely cause of this change 
in transit duration.}


   \maketitle

%


\section{Introduction}

The study of transits is one of the most powerful methods in exploring extrasolar planet properties. 
Although transits events are rare, viewing geometry dependent phenomena, they can provide information about extrasolar planets and that is otherwise inaccessible. Space missions such as CoRoT or the Kepler mission use the transit method in surveys of extrasolar planet systems. The transit light curve incorporates all relevant physical system parameters and, therefore, a highly accurate extrasolar planet transit light curve provides access to this information. The orbits of the planets in our solar-system as well as the Moon and artificial Earth-orbiting satellites are known to undergo secular changes in their orbit parameters.  The reasons for these changes are the gravitational attraction of other ("third") bodies, deviations from spherical symmetry, air drag, non-gravitational forces and relativistic effects among others.  Similarly, the orbits of extrasolar planets are expected to vary at some level, although, there is so far no evidence that the orbit or the physical parameters of any exoplanet have changed. 

Changes in the orbit inclination $i$ caused by a precession of the orbit plane are particular interest.  Transiting planets are ideal for detecting these changes, especially when the transit is "grazing", i.e., when the planet eclipses only the polar regions of its host star.  As a consequence, extrasolar planets with lower inclination and larger impact 
values (but still producing transits) such as the cases of OGLE-56, TrES-2, TrES-3, and TrES-4 are particularly well suited to detecting these orbital changes.

The TrES-2 exoplanet system was discovered in 2006 by the TrES (Trans-atlantic Exoplanet Survey) project (\cite{2006ApJ...651L..61O}).  Using the Mt. Hopkins Observatory 1.2m FWLO, \cite{2007ApJ...664.1185H} performed the first accurate analysis of TrES-2, using three TrES-2 light curves obtained in the fall of 2006. TrES-2 turned out to be a rather typical hot Jupiter exoplanet of mass $M_{p} = 1.198M_{J}$  and radius $R_{p}=1.222R_{J}$, orbiting a G0V star of mass $M_{*}=0.98 M_{Sun}$ and radius  $R_{*} = 1.003R_{Sun}$ in a 2.47 day orbit.  In this {\it Letter}, we present new transit observations of TrES-2 obtained on May and September of 2008, which suggest that the orbit inclination and hence the transit duration of TrES-2 have changed.

\section{Observations and data reduction}

    \begin{figure*}
    \centering
    \includegraphics[width=6.0cm]{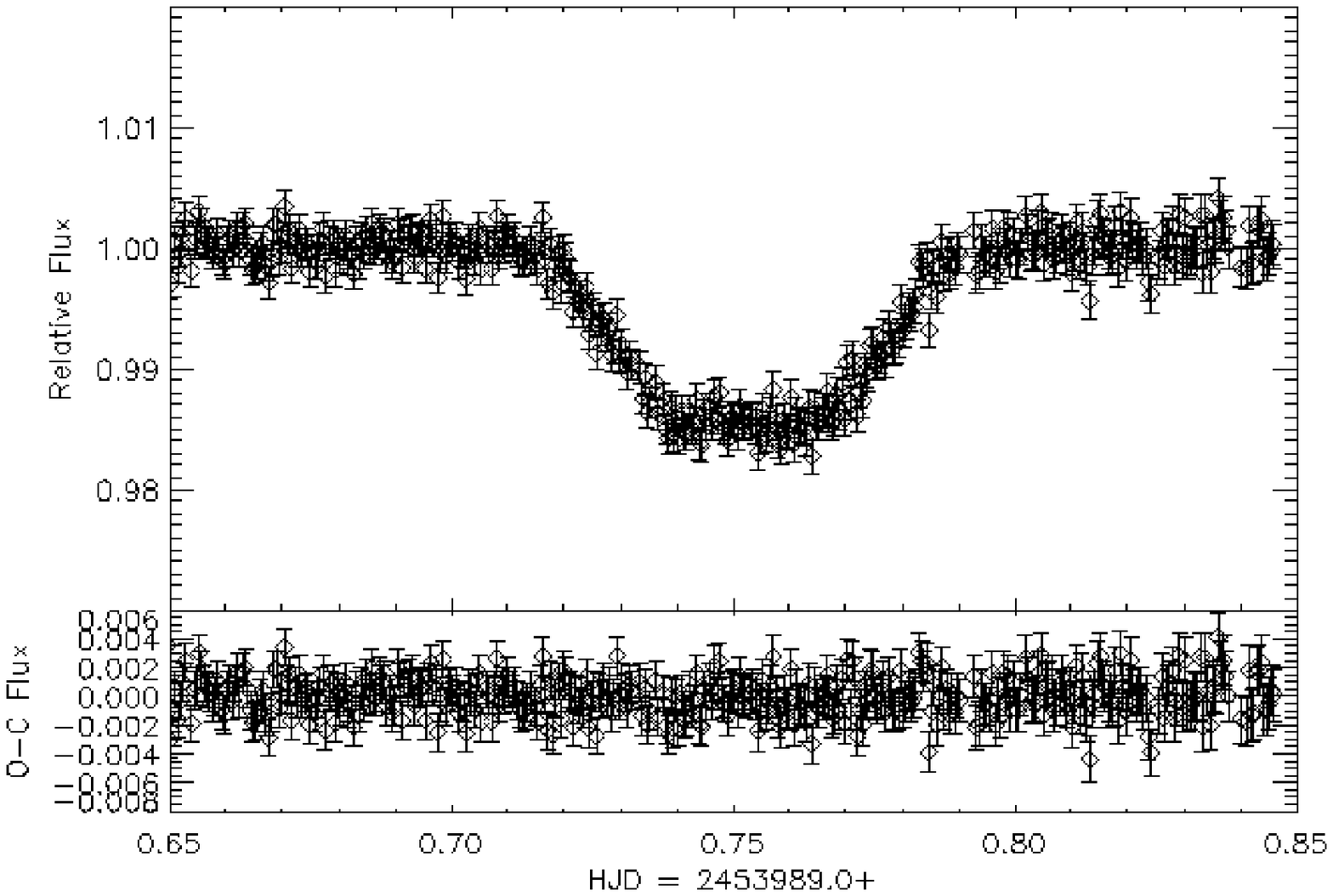}
    \includegraphics[width=6.0cm]{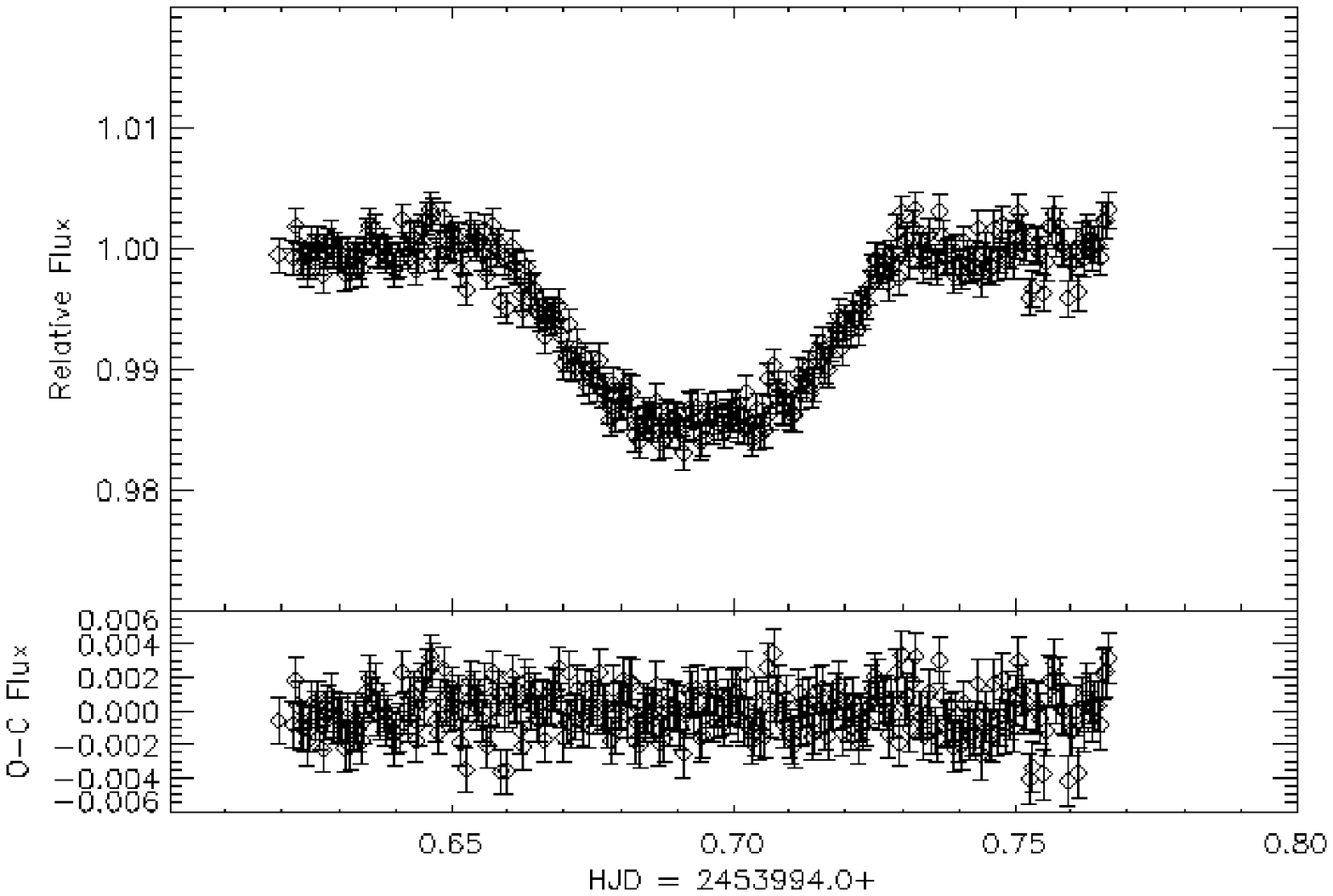}
   \includegraphics[width=6.0cm]{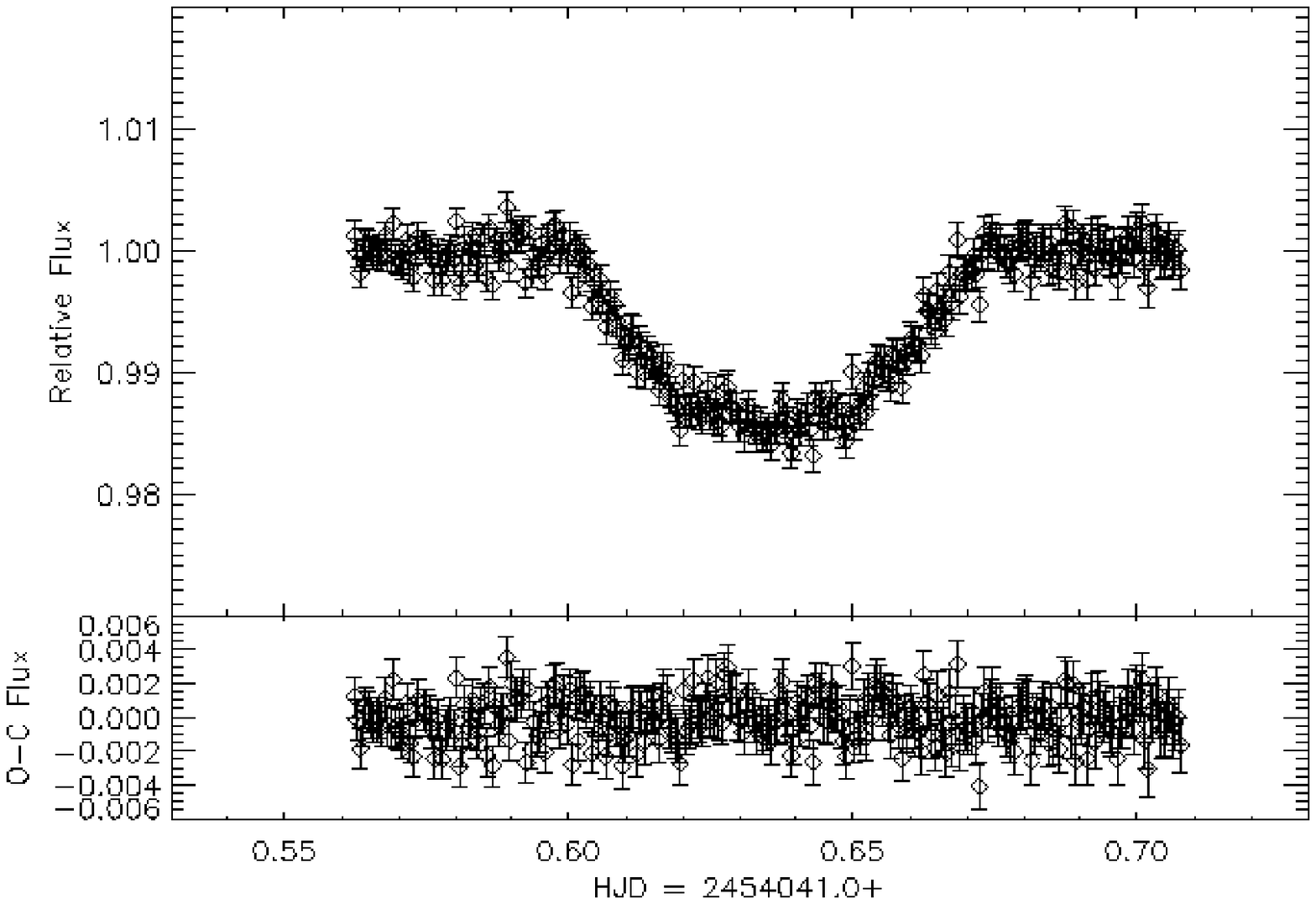}
   \includegraphics[width=6.0cm]{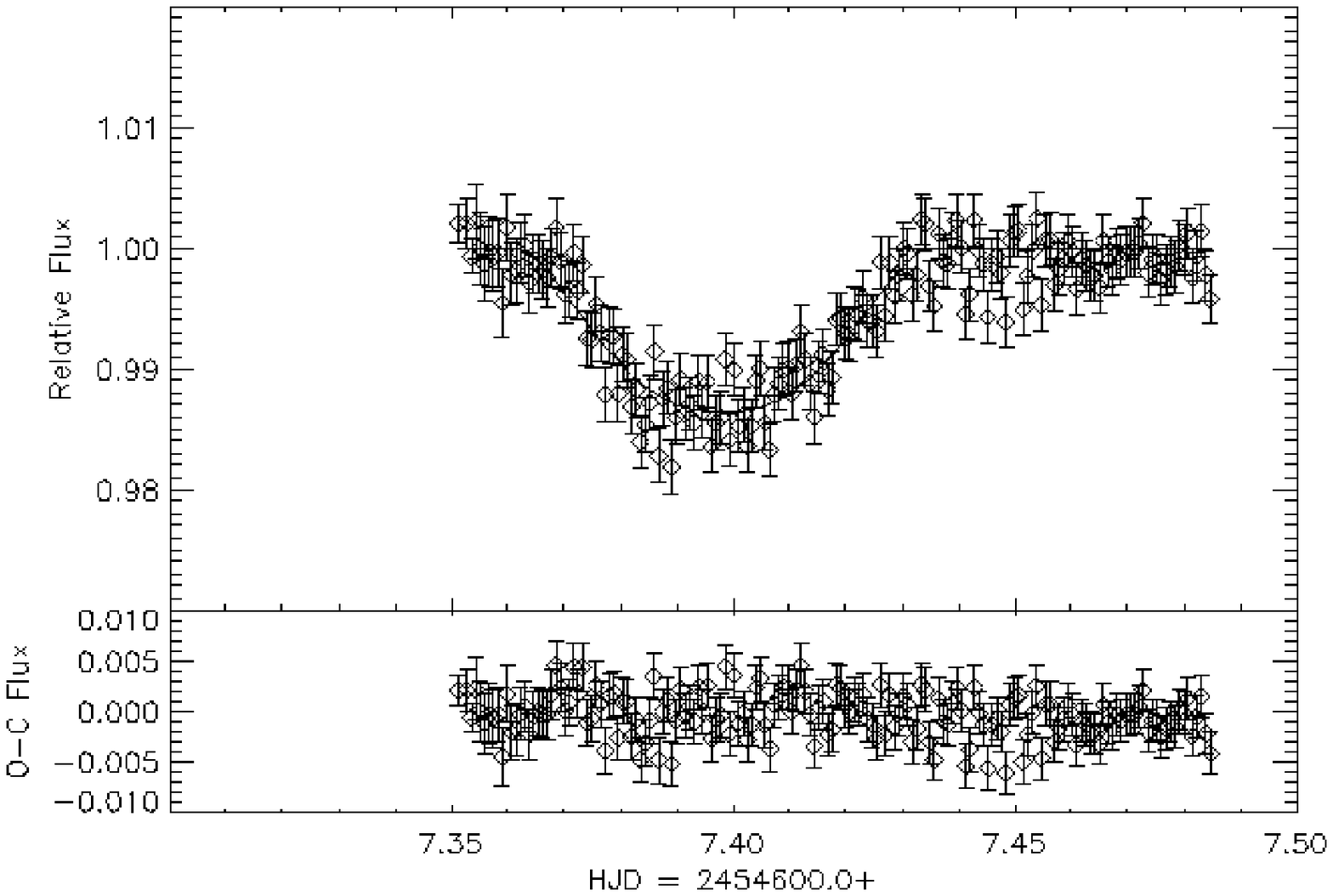}
    \includegraphics[width=6.0cm]{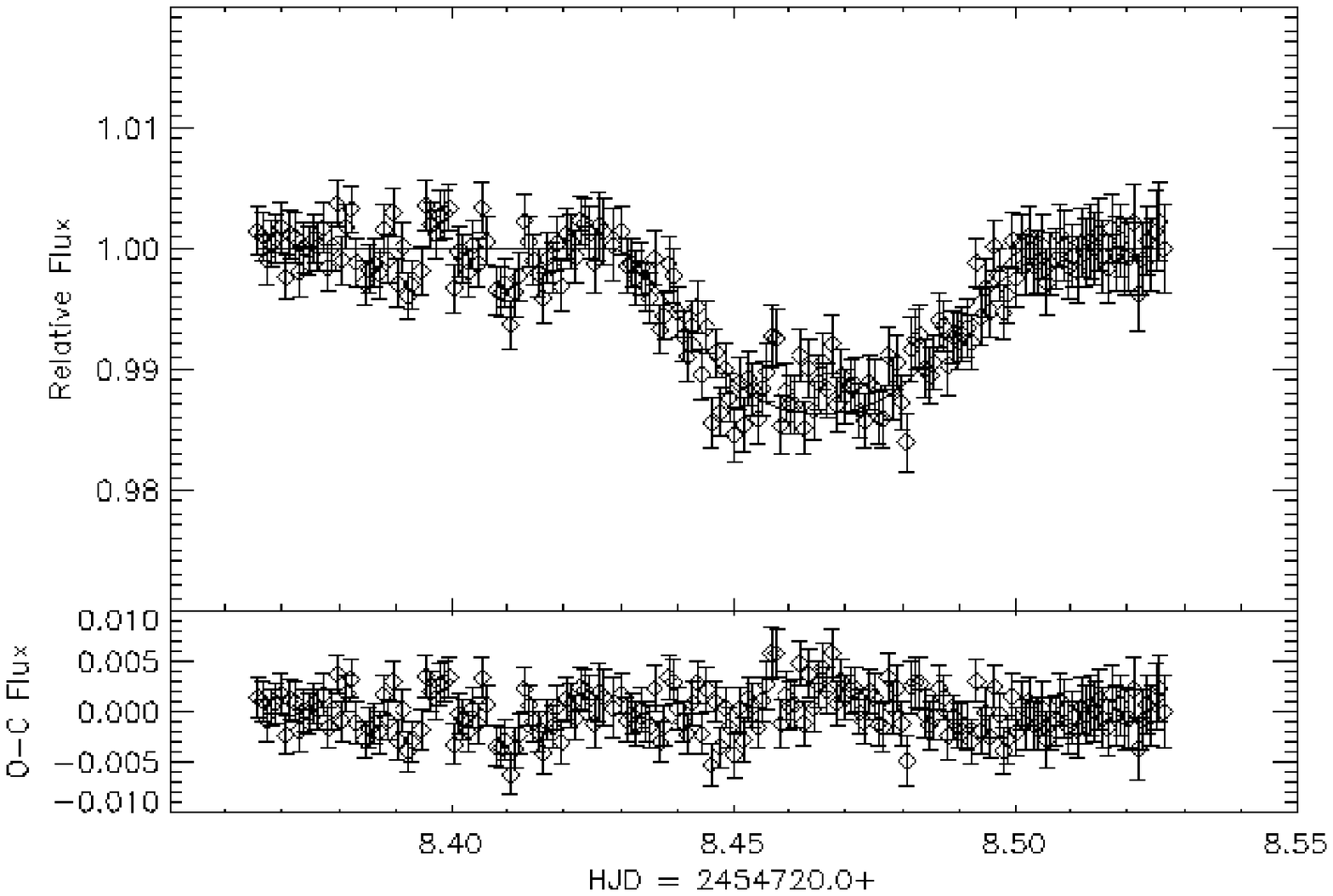}

   \caption{Observed TrES-2 light curves, model fits and residuals (lower panel); upper three light curves were taken 
1.2m FLWO (\cite{2007ApJ...664.1185H}), but fitted with our model code using Holman et al.'s (2007) limb darkening. The 
lower two light curves were taken with the 1.2m OLT at Hamburg Observatory, and fitted with the same model and the appropriate limb-darkening coefficients. The transit duration measured in 2008 is $\sim 3$ minutes shorter than that measured in 2006.}
\label{fig1}
\end{figure*}

We observed two transits of TrES-2 using the ephemeres suggested by \cite{2006ApJ...651L..61O} and by \cite{2007ApJ...664.1185H} from

$T_{c}(E) = 2,453,957.6358 [HJD] + E\cdot(2.47063 \ \mathrm{days})$,
\noindent
using the 1.2m Oskar L\"uhning telescope (OLT) at Hamburg Observatory.  For the first observing run on 20 May 2008, we used a $1Kx1K$ CCD with a $5'x5'$ FOV, readout noise of 4.68 $e^{-}$, and gain of 1.12 $e^{-}/ADU$ without filter (corresponding to a near R band). With a 30-second exposure time and $1x1$ binning, we achieved a time resolution of 1.13 minutes. For the second observing run on 18 September 2008, we used a new CCD camera with a $3Kx3K$ chip and $10'x10'$ FOV, readout noise and gain were 16.37 $e^{-}$ and 1.33 $e^{-}/ADU$ respectively, with an i-band filter. With this setup, we used a 60-second exposure time and a binning of $2x2$, which provided an effective time resolution of 1.17 minutes.  During our observations, the airmass values ranged from 1.661 to 1.081 and from 1.0423 to 1.7176, during the first and second observing runs, respectively, and the seeing was typically 2.93" and 1.85", respectively, which are quite typical of the Hamburg site.

For the data reduction, we used \textit{Starlink} and \textit{DaoPhot} software routines, and the \textit{MATCH} code. After applying bias subtraction, dark correction, and flat fielding, we continued with our photometry. For TrES-2, we selected the aperture photometry mode using apertures centered on the target star, check stars, and sky background. Typical sky brightness values (the night sky is quite bright near Hamburg) for the first and second night was 325.6 and 270.2 ADUs, respectively, i.e., values at a level $1.5\%$ and $1.25\%$ of the star's flux, respectively.  We used a total of 4 reference stars to test and calibrate the light curve $(U1320+10215660 - 10214928 - 10220179 - 10219455)$. To estimate the magnitude errors, we followed \cite{2001phot.work....1H} and used the expression

\begin{equation}
\sigma_{mag} = 1.8057 \displaystyle \frac{\sqrt{N_{*}+p}}{N_{*}}, \label{eq1}
\end{equation}

\noindent
where $p=n_{pix}(1+\frac{N_{pix}}{n_{B}})(N_{S}+N_{D}+N_{R}^{2}+G^{2}\sigma_{f}^{2})$, $N_{*}$ is the total number of photons, $N_{pix}$ is the number of pixels, which was used to define $N_{*}$, $n_{B}$ is the total number of background pixels, $N_{S}$ is the number of total number of photons per pixel from background, $N_{D}$ is the total number of dark electrons per pixel, $N_{R}^{2}$ is the total number of electrons per pixel due to the readout noise, $G$ is the gain, and $\sigma_{f}^{2}$ is the 1$\sigma$ error of the A/D converter ($\sim$ 0.289). The factor of 1.8057 converts the errors in flux (electrons) into errors in magnitudes (\cite{2001phot.work....1H}), we made no further changes to the adopted values of the measurement errors and continued with the light curve analysis with transit model fitting.  Our final relative
photometry is presented in Table.\ref{tab1}, which is available in its entirety in machine-readable form in the electronic version of this paper.

  \begin{table}

      \caption[]{Relative Photometry data.}

         \label{tab1}

         $\begin{array}{lll}

            \hline

            \noalign{\smallskip}

             HJD     &  Relative Flux & Uncertainty \\

            \noalign{\smallskip}

            \hline

            \noalign{\smallskip}

              2454607.3512 & 1.001980 & 0.0015717      \\
              2454607.3528 & 1.002020 & 0.0021024      \\
              2454607.3536 & 0.999300 & 0.0014341      \\
              2454607.3543 & 1.002150 & 0.0031083      \\
              2454607.3551 & 0.999920 & 0.0030757      \\
              2454607.3559 & 0.998690 & 0.0030854      \\

            \noalign{\smallskip}

            \hline

         \end{array}$

Note : Table 1 is available electronically. The time stamps refer to the Heliocentric Julian Date (HJD) at the end of each exposure.
   \end{table}

\section{Model analysis}

There are various ways of creating a physical transit model as described by \cite{2002ApJ...580L.171M}, \cite{2008arXiv0807.4929W}, or \cite{2008MNRAS.389.1383K}. The main physical parameters of an exoplanet system are the host star mass and radius $M_{*}$, $R_{*}$, the planet mass and radius $M_{p}$ and $R_{p}$, the orbit semimajor axis $\alpha_{orbit}$, the orbit inclination  $i_{orbit}$, and the limb darkening law.
There are also indirect or secondary parameters that can be computed once the main parameters are known, or which have no specific physical meaning or cannot otherwise be measured. These parameters include the central transit time $T_{c}$, which can be observed and used as a phase reference.  The planetary period $P$ must be determined by other means (i.e., not from the shape of the transit light curves), the eclipse duration and depths are functions of the orbit inclination, the stellar and planetary sizes and the limb darkening law.  In our modelling, we assume exactly spherical stars and planets with radii independent of the observing wavelength.  We also assume circular orbits and -- of course -- the validity of Kepler's third law, which links the parameters $\alpha_{orbit}$, $M_{*}$, and $M_{p}$, since the period $P$ is known. Usually one needs to fit all the main light curve parameters simultaneously, although, we clearly demand that the masses of star and planet, their radii, and the semi-major axis of the orbit remain constant for all our transit observations. 

Following \cite{2002ApJ...580L.171M} and \cite{2007ApJ...664.1185H}, we assume a quadratic limb darkening law of
\begin{equation}
\frac{I_{\mu}}{I_{o}} = 1 - u_{1}(1-\mu) - u_{2}(1-\mu)^{2}, \label{eq2}
\end{equation}
where $\mu$ denotes the cosine of the angle between the surface normal and the observer, and $I_o$ and  $I_{\mu}$ are the intensities at disk center and angle $\mu$ respectively; a complete list of the system parameters is given in Table \ref{tab2}. While in principle the limb darkening coefficients could be fitted by a transit light curve, limb darkening is indicated by the structure of the star's atmosphere and should not have to be adjusted to fit a specific transit light curve.
Based on \cite{2004A&A...428.1001C} and \cite{2007ApJ...664.1190S}, we used the values $u_{1}$ = 0.318 and $u_{2}$ = 0.295 for the i filter and $u_{1}$ = 0.430 and $u_{2}$ = 0.265 for the R filter.   The relation between duration and inclination is given by Eq. \ref{eq3} (\cite{2006ApJ...636..445C}; \cite{2003ApJ...585.1038S})

\begin{equation}
D = \displaystyle\frac{P}{\pi}\arcsin \left[  \frac{R_{*}}{\alpha \sin i} \sqrt{\left( 1+\frac{R_{p}}{R_{*}} \right)^2-\left( \frac{\alpha\cos i}{R_{*}} \right)^2}  \right], \label{eq3}
\end{equation}
\noindent
where all parameters have the meanings explained above and $D$ denotes the total transit duration. It is important to realize that the duration of a transit as defined by Eq. \ref{eq3} is not a function of the assumed limb darkening coefficients (see also \cite{2007ApJ...655..564K}), since the stellar and planetary radii are assumed to be wavelength independent;  only the transit shape and depth, but not the transit duration depend on the limb dark coefficients. 

  \begin{table}

      \caption[]{TrES-2 physical parameters (adopted from Holman et al.)}

         \label{tab2}

         $\begin{array}{p{0.5\linewidth}l}

            \hline

            \noalign{\smallskip}

            TrES-2 system parameters      &  Values \\

            \noalign{\smallskip}

            \hline

            \noalign{\smallskip}

            $M_{*}$   &$ 0.98 $M_{o}     \\

            $M_{p}$   & $ 1.198 $M_{J} \\

            $R_{*}$   & $ 1.003 $R_{o} \\

            $R_{p}$   & $ 1.222 $R_{J} \\

            Orbit Radius   &$ 0.0367 $AU     \\

            Period   & $2.470621 $days \\

            \noalign{\smallskip}

            \hline

         \end{array}$

   \end{table}


To determine the best fit model parameters, we used the $\chi^{2}$ statistic of

\begin{center}

\begin{equation}
\chi^{2} = \displaystyle\sum_{j=1}^{N_{F}} \left[ \frac{F_{j,obs} - F_{j,mod}}{\sigma_{j}} \right]^{2} , \label{eq4}
\end{equation}

\end{center}

\noindent
where $N_{F}$ denotes the number of available data points, $F_{j,obs}$ the flux, $F_{j,mod}$ the model flux, and $\sigma_{j}$ the error in $F_{j,obs}$ all at time j. To ensure a consistent analysis, we re-analyzed the light curves already presented by \cite{2007ApJ...664.1185H} using the appropriate limb darkening coefficients in addition to our own new transit observations.


To determine the error in the derived fit parameters, we used the following bootstrap procedure, which has some
similarities to a Markov chain error analysis. In step (1), we obtained the best-fit light curve minimizing Eq. \ref{eq4}. In step (2), we computed the best-fit model residuals $FR$ defined by $(FR_{j} = F_{j,obs}-F_{j,mod})$.  In step (3), we determined the second moment $\sigma$ of the residual distribution, assuming a Gaussian distribution. In step (4), we generate a random light curve $(RL_{j})$ using the model and the derived value of $\sigma$, and finally in step (5) determined a best-fit to this randomized light curve by repeating step (1). This procedure was repeated typically 1000 times and we recorded the variations in the fit parameters such as inclinations and durations.

   \begin{table}

      \caption[]{Duration, inclination, $\chi^{2}$ values and limb darkening coefficients from five light curve fits; units for duration and errors are minutes and for inclination and errors are degrees.}

         \label{tab3}


         $\begin{array}{llllllll}

            \hline

            \noalign{\smallskip}

            \textit{Tc}$ $time$ $[HJD] &  Duration & Errors & Inclin. & Errors & \chi^{2} value & LDL \\

            \noalign{\smallskip}

             \hline

            \noalign{\smallskip}

            3989.7529\pm0.0069  & 110.308 & 0.432 &  83.59  &  0.019 &  432.1 & S1 \\

            3994.6939\pm0.0066  & 109.230 & 0.448 &  83.56  &  0.019 &  296.8 & S1\\

            4041.6358\pm0.0070  & 109.025 & 0.430 &  83.55  &  0.019 &  290.6 & S1\\

            4607.4036\pm0.0072  & 106.620 & 0.883 &  83.44  &  0.036 &  179.1 & S2\\

            4728.4740\pm0.0071  & 106.112 & 0.870 &  83.43  &  0.036 &  190.1 & S3\\

            \hline

            3989.75286   & 113.275 &        &  83.74  &        &  443.0 & S2 \\

            3989.75286   & 111.632 &        &  83.66  &        &  435.6 & S3 \\

            3994.69393   & 111.450 &        &  83.65  &        &  305.5 & S2\\

            3994.69393   & 110.195 &        &  83.60  &        &  300.1 & S3\\

            4041.63579   & 111.790 &        &  83.67  &        &  296.6 & S2\\

            4041.63579   & 110.197 &        &  83.60  &        &  292.4 & S3\\

            4607.40356   & 104.317 &        &  83.35  &        &  180.0 & S1\\

            4607.40356   & 105.060 &        &  83.38  &        &  179.1 & S3\\

            4728.47400   & 104.890 &        &  83.37  &        &  195.7 & S1\\

            4728.47400   & 107.100 &        &  83.47  &        &  192.6 & S2\\

            \noalign{\smallskip}

            \hline\hline

         \end{array}$

   \end{table}

\section{Results}


With the method described above we analyzed all available transit observations (cf., Fig. \ref{fig1}); the results of our fits are summarized in Table \ref{tab3}, where we also indicate the darkening coefficients used in the analysis (upper five entries).
For the Table \ref{tab3}, we used three sets of limb dark coefficients. Set-1 for Holman's data (S1:$u_{1}=0.22$, $u_{2}=0.32$), Set-2 for OLT's data in R filter (S2:$u_{1}=0.430$, $u_{2}=0.265$), and finally, Set-3 for OLT's data in i filter (S3:$u_{1}=0.318$, $u_{2}=0.295$). In the first part of table, we listed the best-fit results using the most suitable set of limb dark coefficients. In the second part, we used all the possible combinations of limb dark coefficients to show that duration is not strongly affected by limb darkening. Holman's light curves have 433, 301, and 299 data points, and ours have 169 and 193 data points, respectively. With our new observations, we now have precise measurements of $T_{c}$ for five transits, so we are able to refine the ephemeres equation. A least squares fit to these five transit events results in the expression \\
$T_{c}(E) = 2,453,957.63403\,\mathrm{[HJD]} + E\cdot(2.4706265 \ \mathrm{days}).$  

\begin{figure}[h]
\centering

\includegraphics[width=8cm,angle=0,clip=true]{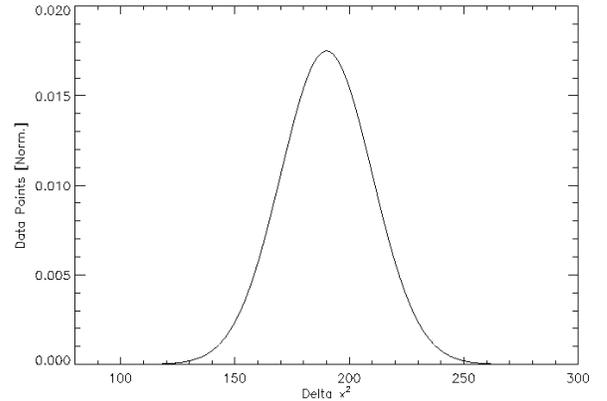}

\caption{Fit improvement $\Delta \chi^2$ for 1000 Monte Carlo realisations when going from models of fixed inclination to models of free inclination (see text for details).}

\label{fig2}

\end{figure}


In our basic analysis, we assumed the period, masses, radii, and semi-major axis of the orbit to have the values derived by \cite{2007ApJ...664.1185H}.  With this approach, we found that the OLT data taken in 2008  appear to prefer different inclinations than the FLWO data taken in 2006 (cf., Table \ref{tab3}).  To assess the significance of these inclination differences, we carried out the following analysis. We assumed that the five transit light curves (cf., Fig. \ref{fig1}) can be described  by variable stellar and planetary radii, variable inclination, and variable central transit times (model A; 20 fit parameters), and generated 1000 realizations of this light curve with the method described above. In model B (10 fit parameters), we fixed the stellar and planetary radii to a common value, but still allowed the inclinations and central transit times to vary for each data set, while in model C with (5 fit parameters), we also kept all inclinations constant at the best-fit model value and allowed only the transit times $T_c$ to vary. The generated random light curves were then fitted (via $\chi^2$ minimisation) by models (A), (B),  and (C), and the errors were computed as described above; clearly, the closest fits are expected for model (A), and the poorest fits for model (C), which is, however, the "canonical" model.  The essential question is whether the fit improvement between models C, B, and A is significant or not ? \par
We first consider the fit improvement between models (C) and (B); for each generated light curve, we can compute the appropriate $\Delta \chi^2$, and the distribution of these $\Delta \chi^2$ residuals is shown in Fig. \ref{fig2}.  A typical value of $\Delta \chi^2$ is clearly $\approx$ 200, and an F-test indicates that the fit improvement between model (C) and model (B) is highly significant; carrying out the same exercise for going from model (B) to model (A), we find any additional fit improvement is insignificant.  This does of course meet our physical expectations, since the sizes of the star and planet are not expected to change, while the orbit inclinations might change.

\begin{figure}[h]

\centering

\includegraphics[width=8cm,angle=0,clip=true]{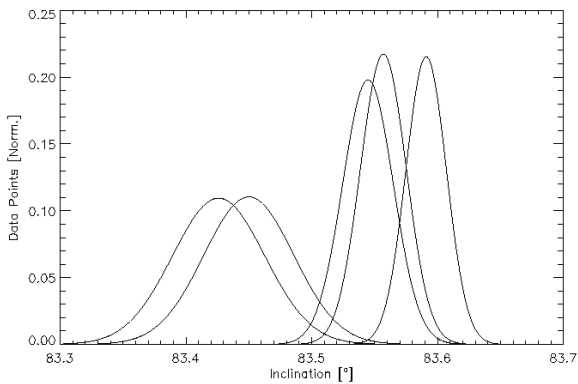}

\caption{Inclination error distribution derived from 1000 $Monte$ $Carlo$ bootstraps: OLT light curves (left) and FLWO light curves (right).}

\label{fig3}

\end{figure}

The bootstapped error distribution of the derived inclinations in model (B) at the five epochs is shown in Fig. \ref{fig3}, which shows the distributions to have very little overlap. We thus conclude that from a statistical point of view the change in inclination observed in the 2008 data is very significant.

Can the apparent change in inclination be caused by errors in the analysis?  We investigated possible non-physical causes of the derived inclination change. Our own OLT observations were not carried out with the same filter as the observations by \citet{2007ApJ...664.1185H}. Since our fit approach explicitly assumes that the stellar radius is independent of wavelength, the eclipse duration defined by Eq. \ref{eq3} depends only on both the sizes of the host star and planet and the orbit inclination, but not on the shape of the transit light curve, which depends on the chosen limb darkening law.  Since we keep the stellar and planetary radius fixed in our fitting scheme (model B), a change in inclination is the only way to produce a change in eclipse duration. However, since the ontire observed light curve is fitted to an assumed model light curve, we can obtain an incorrect fit and thus an incorrect inclination and duration if we assume an inappropriate limb darkening law.


Since the Hamburg OLT observations were carried out in "bluer" filters (than the filter which Holman used) with larger contrasts between disk center and limb, we assessed the influence of limb darkening by analyzing all data for all limb darkening coefficients, assuming a quadratic limb darkening law specified by Eq. \ref{eq2}; the results of these fits are also given in Table \ref{tab3} (lower ten entries).  As is clear from Table \ref{tab3}, the derived durations depend on the assumed limb darkening coefficients, although, the inclination discrepancy between the 2006 and 2008 data cannot be attributed to limb darkening.  We further note that the nominal limb darkening coefficients provide the best fits in all cases. 

Finally, we asked ourselves which sets of limb darkening model coefficients would provide the best-fit light curve to our new OLT data, keeping the transit duration fixed at the values observed in 2006. With this fit setup we found poorer fits ($\chi^{2}=204.58$) with unrealistic limb darkening coefficients of $u_{1}=0.4258$ and $u_{2}=0.2254$, corresponding to $T_{eff} \sim 4800^{o}K$, for epoch-312 (\cite{2004A&A...428.1001C}), too low given the spectral type of TrES-2's host star. We repeated the experiment for all combinations of limb darkening coefficients in the range $0.00 < u_1+u_2 < 1.00$, and found a best-fit model with $\chi^{2}=182.0$, although, the corresponding coefficients of $u_1=1.19$ and $u_2=-0.80$ do not relate to any significant physical meaning. We therefore conclude that the discrepancy between the derived durations of the 2006 and 2008 transits cannot be explained by the assumed limb darkening laws.

Another possible source of error is an incorrect determination of the out-of-transit flux. All assumed model light curves assume an out-of-transit flux of exactly unity.  The data points are normalized to unity by the available out-of-transit data points, and therefore the accuracy of this normalization depends on the number of such out-of-transit data points and their respective errors; we note that in the extreme case of zero out-of-transit data points no normalization to unity can be carried out and therefore no meaningful analysis.  If the data normalization were for some reason too low, the observed transit depth would be deeper and hence the derived inclination higher (and duration longer); what must then concern us is a data normalization that is too high, reducing the eclipse depth and hence reducing inclination and duration.  To study the influence of errors in the data normalization, we introduced a multiplicative amplitude factor to the model, which should nominally be unity but which was allowed to vary to obtain a best fit. In all cases, we found that these amplitudes were close to unity (at the level of a few times 10$^{-4}$) without any substantial change in transit duration. We thus conclude that normalization errors do not significantly affect our results. 

   \begin{figure}[h]

   \centering

   \includegraphics[width=8cm,angle=0,clip=true]{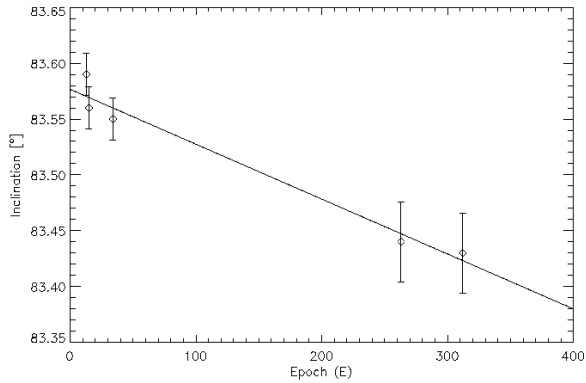}

\caption{Epoch versus inclination: a linear fit is clearly sufficient to describe the available data; if the inclination continues to decrease, the minimum inclination threshold $(i_{min}=81.52^{o})$ will be reached around epoch $\sim$ 4179.}

              \label{fig4}

    \end{figure}

Finally, following \cite{2007ApJ...664.1185H}, our modelling approach assumes a spotless star. If the true star were spotted, our modelling approach and thus the inferred model parameters would be incorrect. Clearly, if one were to place spots exactly at those limb positions where the planets enters and leaves the stellar disk, the apparent transit duration would change; at the same time the limb darkening and transit depth would change and it is unclear, what the resulting fit parameters of such an inappropriate model would be.  Since we, first,  have two transit observations yielding consistent fit results and, second, the $v sin(i)$ of TrES-2's host star is quite low $(<2.0$ $km/sec$, \cite{2006ApJ...651L..61O}) proposed that it was an inactive star and we conclude that star spots are an unlikely explanation of the observed transit duration shortening.

If we accept that the formally measured inclination changes are real, we can study (in Fig. \ref{fig4}) the variation in inclination versus time derived from the presently available data; clearly, a linear fit is sufficient for a description. In every transit systems, two theoretical thresholds of minimum inclination angles exist, a first threshold, where the planet is never fully in front of the stellar disk, and a second threshold, where no transit occurs at all.  For the TrES-2 system parameters listed in Table \ref{tab2}, the first inclination threshold occurs for $i_{min,1}=83.417^{o}$, and the second threshold for $i_{min,2} = 81.52^{o}$.  From our linear fit shown in Fig. \ref{fig4}, the first threshold is reached around $Epoch \sim 325$ corresponding to October 2008 with an eclipse duration of $D=105.92$ minutes.

\section{Conclusions}

Because its orbital properties of TrES-2 is a particularly interesting exoplanet. The transits of TrES-2 are almost "grazing" and are therefore quite sensitive to inclination changes. By comparing our new 2008 observations of TrES-2 with data obtained in 2006, we have detected an inclination change of 0.1 degress in a little under two years, corresponding to a transit duration change of about $\sim 3.16$ minutes. We have investigated the influences of limb darkening and the light curve normalization on the derived inclinations and found that the observed duration changes cannot be attributed to incorrect light curve modelling. We had therefor to search for a physical cause of the observed changes. Since stellar and planetary radii are unlikely to have changed and an explanation related to star spots appears unlikely given the activity status of TrES-2's host star, a change in the orbital inclination is the only remaining explanation. A third body in the form of an outer planet, which possibly does not cause any transit eclipses, would provide the most straightforward explanation. We emphasize that it  
appears extremely worthwhile to continue transit monitoring of TrES-2 in the 2008 - 2010 time frame to study its inclination changes as well as to search for further companions of TrES-2 in radial velocity data.  In the Kepler field, high precision photometry of hundreds of transits of TrES-2 will be obtained, which should allow a precise determination of all orbit changes as well as a refinement of the stellar and planetary parameters of the TrES-2 exoplanet system.

\begin{acknowledgements}

We want to express our gratitude to Dr. Hans-J\"{u}rgen Hagen for his superb technical support at the OLT. We also want to thank Drs. R. Wichmann and M. Esposito and Prof. Dr. G. Wiedemann for useful discussions.  DM was supported in the framework of the DFG-funded Research Training Group ''Extrasolar Planets and their Host Stars'' (DFG 1351/1).

\end{acknowledgements}

\bibliographystyle{aa}

\bibliography{aa}

\end{document}